\documentclass[twocolumn]{aastex631}
\usepackage[utf8]{inputenc}

\def\gtrsim{\lower 2pt \hbox{$\, \buildrel {\scriptstyle >}\over
{\scriptstyle \sim}\,$}}
\def\lesssim{\lower 2pt \hbox{$\, \buildrel {\scriptstyle <}\over
{\scriptstyle \sim}\,$}}

\def\hst{{\sl HST}}

\def\chandra{{\sl Chandra}}

\def\hst{{\sl HST}}

\newcommand{\as}{$^{\prime\prime}$}

\def\approxlt{\lower.2em\hbox{$\buildrel < \over \sim$}}
\def\approxgt{\lower.2em\hbox{$\buildrel > \over \sim$}}

\def \xs{\hbox{CXOUJ114106.0+322515}}
\def \sou{\hbox{KUG 1138+327}}

\begin{document}

\title{Serendipitous discovery of a young cluster of galaxies at $z \sim 0.5$\\
projected next to the nearby tadpole galaxy KUG 1138 + 327}

\author[0000-0002-9279-4041]{Q. Daniel Wang} \thanks{Contact e-mail:wqd@umass.edu} 
\affiliation{Department of Astronomy, University of Massachusetts, Amherst, MA 01003, USA}

\author{Juergen Ott}
\affiliation{National Radio Astronomy Observatory. 1011 Lopezville Road. Socorro, NM 87801, USA}

\begin{abstract}

Using a 90 ks Chandra ACIS-S observation in the 0.3–7 keV band, along with complementary Low-Frequency Array and Karl G. Jansky Very Large Array data in the 120–168 MHz and 1–2 GHz ranges, we study the diffuse emission around the nearby dwarf galaxy KUG 1138+327. Our analysis reveals a diffuse X-ray feature on the southern side, disconnected from the galactic disk. This feature exhibits a hard X-ray spectrum, which is highly unusual for an outflow from a dwarf galaxy. We interpret the irregularly shaped feature as hot plasma in a young galaxy cluster at redshift 0.5, supported by X-ray spectral fitting and consistent with the optical redshift of the central elliptical galaxy of a known cluster identified by a red sequence. Additionally, we detect a radio lobe east of the X-ray feature, likely produced by an AGN offset from the cluster center and confined primarily by ram pressure from the ambient medium. The lobe shows a steep nonthermal radio spectrum,  suggesting a cosmic-ray age of $\gtrsim 5 \times 10^7$ years. Assuming energy equipartition between cosmic rays and magnetic fields, we estimate the lobe's total energy to be $\sim 9 \times 10^{56} {\rm~erg}$, comparable to the thermal energy in the same volume. This study thus identifies a background young cluster projected next to KUG 1138+327 and highlights the potentially significant role of off-center AGN feedback in shaping the intracluster medium.
\end{abstract}


\section{Introduction} \label{s:intro}

Feedback from active galactic nuclei (AGNs) is expected to play an essential role in regulating the circumgalactic and intracluster medium. For example, a recent theoretical study, based primarily on order-of-magnitude estimates, shows that cosmic rays (CR) produced by a central AGN may be a key source of feedback in $\sim 10^{13-14} {\rm~M_\odot}$ halos \citep{Quataert2025}. Particularly important is the feedback during the primary epoch of supermassive black hole growth at $z \sim 1 -3$. But even at lower redshifts, CR feedback can be significant, preferentially on the outskirts of such halos. Specifically, the CR pressure can be comparable to or exceed the thermal gas pressure at radii of the order of the virial radius. However, the existing study is limited to feedback from the central galaxies of the halos \citep{Quataert2025}, which may have significantly underestimated the overall role of CR.  The in situ energy injection by satellite galaxies at off-center radii of the halos, especially in their early assembling stages,  could make this feedback scenario even more appealing. 

\begin{figure}[!htb]
\centerline{
\includegraphics[width=1\linewidth,angle=0]{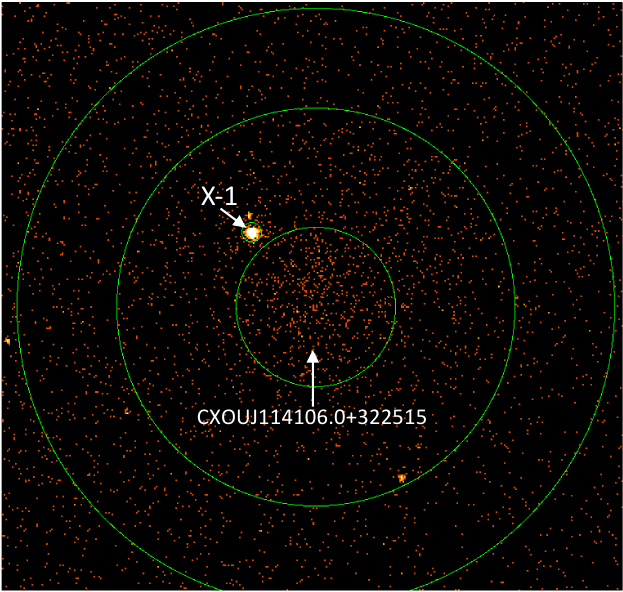}}
\caption{
\chandra\ ACIS-S count image of the field around \sou. The inner circle of radius 0\farcm4, centered at R.A. and Dec (J2000) $= 11^h41^m6^s,+32^\circ25^{\prime}15^{\prime\prime}$, outlines the region where the on-source spectrum of the diffuse feature (CXOUJ114106.0+322515) is extracted, while the concentric annulus with the inner and outer radii of $1^{\prime}$ and 1\farcm5  marks the region where the local background is estimated.  
 }
\label{f:f2}
\end{figure}

Here we present a serendipitous discovery of a young cluster at $z \sim 0.5$. It contains a diffuse hot plasma not completely virialized and an off-center radio lobe emanating from a compact optical object (presumably an AGN), probably falling towards a young cluster core.  This phenomenon demonstrates that off-center CR injection can be an important process in galaxy cluster formation.
\begin{figure}[!htb]
\centerline{
\includegraphics[width=1\linewidth,angle=0]{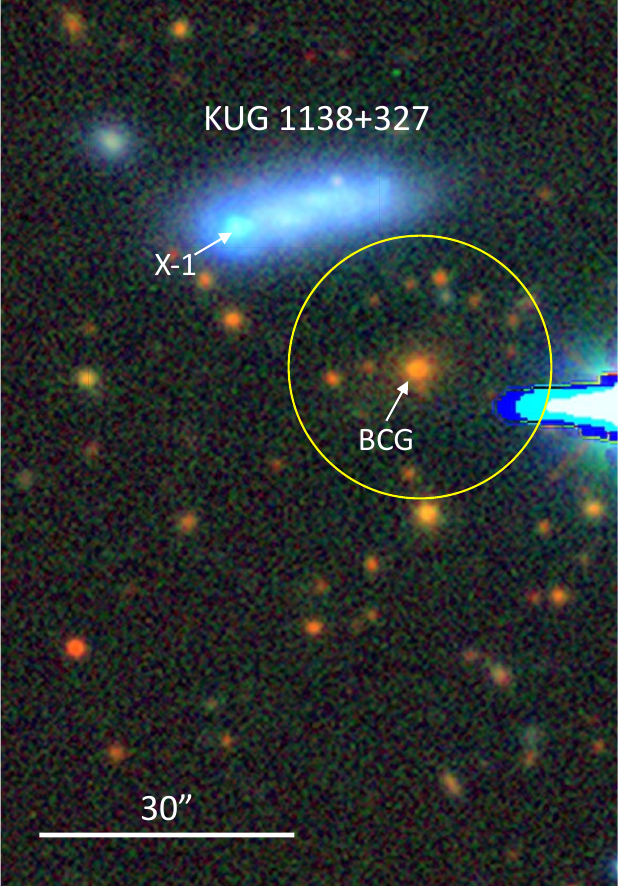}}
\caption{
Dark Energy Spectroscopic Instrument (DESI) Legacy Survey DR9 image, showing a concentration of red galaxies south of \sou. The g, r, and z filters of the survey are mapped to the blue, green, and red channels in the color image, obtained from https://www.legacysurvey.org/viewer.  A concentration of early-type red galaxies (orange colored here) is apparent in the encircled field around the bright central galaxy (BCG), roughly at the morphological center of the diffuse X-ray feature (see Fig.~\ref{f:f2}), and represents a galaxy cluster cataloged by \cite{Hao2010}. Also marked is the location of the X-ray source, X-1, in the starburst region of \sou.
 }
\label{f:f1}
\end{figure}
The discovery of the young cluster is based chiefly on a \chandra\ observation targeted at the dwarf galaxy, KUG 1138+327 (also known as Kiso\,5639),  at a distance of 24.5 Mpc; \citep[Fig.~\ref{f:f1}; 1\as\ = 119~pc][]{Wang2025}.  This remarkable dwarf galaxy has a tadpole appearance that arises from an extraordinary starburst region at one end of the galactic disk \citep[e.g.,][]{Elmegreen2016}. 
We have reported the discovery of an extreme ultraluminous X-ray source (ULX), X-1 (Figs.~\ref{f:f2} and \ref{f:f3}), which appears to be associated with the central stellar cluster of the starburst. 
The observation further indicates the presence of an intriguing diffuse X-ray emission enhancement south of the galaxy, which was initially suspected to be due to the outflow from the starburst. This enhancement in X-ray emission may also be associated with an extended radio feature. This paper presents results on these diffuse radio and X-ray emissions. 

The remainder of the paper is organized as follows. We describe the observations and data reduction in \S~2, present the data analysis and results in  \S~3, discuss their implications in \S~4, and summarize our main findings in \S~5.  Throughout this paper, we assume the $\Lambda$CDM cosmology\footnote{\url{https://astro.ucla.edu/~wright/CosmoCalc.html}} 
with {H\textsubscript{o}} = 69.6 ${\rm~km~s^{-1}~Mpc^{-1}}$, {$\Omega$\textsubscript{M}} = 0.286 and {$\Omega$\textsubscript{$\Lambda$}} = 0.714.

\section{Observations and Data Reduction}\label{s:obs}

\subsection{X-ray data}
The basic reduction of the X-ray observation with a total exposure of 95 ks has been described in \cite{Wang2025}, including source detection.   For the present study, we produce count images in the 0.3-0.7, 0.7-1.5, 1.5-3, and 3-7 keV bands. The corresponding exposure images are constructed with the weight assuming a power law of a photon index of 1.7. They are corrected for bad pixels, as well as photon-energy-dependent telescope vignetting and quantum efficiency variation, which is particularly important at low energies ($\lesssim 1.5$ keV). Dividing the count images by the exposure images gives the intensity images, which can be combined to form an intensity image in a broader band (e.g., 0.3-7 keV). To study diffuse X-ray emission, we remove a region of twice the 70\% energy encircled radius around each source in the count and exposure images. For visual presentation, we further smooth them adaptively to achieve a uniform signal-to-noise ratio across the field. The holes left from the source removal are automatically filled in the resultant diffuse X-ray intensity image.  For simplicity, we use only the S3 chip data for spectral analysis, which uses the software package \texttt{Xspec}  \citep[version 12.14.0; ][]{Arnaud1996}. 

\subsection{Radio data}
Our X-ray analysis is complemented by two radio data sets. The A-array L-band continuum observation was taken with the Karl G. Jansky Very Large Array (VLA) for one hour on 4 July 2023 and had a spatial resolution of 1.3\as (FWHM), while the Low-Frequency Array (LoFAR) Two-metre Sky Survey covered the 120-168 MHz range \citep{Shimwell2017} and had a resolution of 6\as. The combination of these two data sets allows us to characterize the spectral shape of radio emission.

\section{Data analysis and Results}\label{s:res}

\begin{figure*}[!htb]
\centerline{
\includegraphics[width=1.0\linewidth,angle=0]{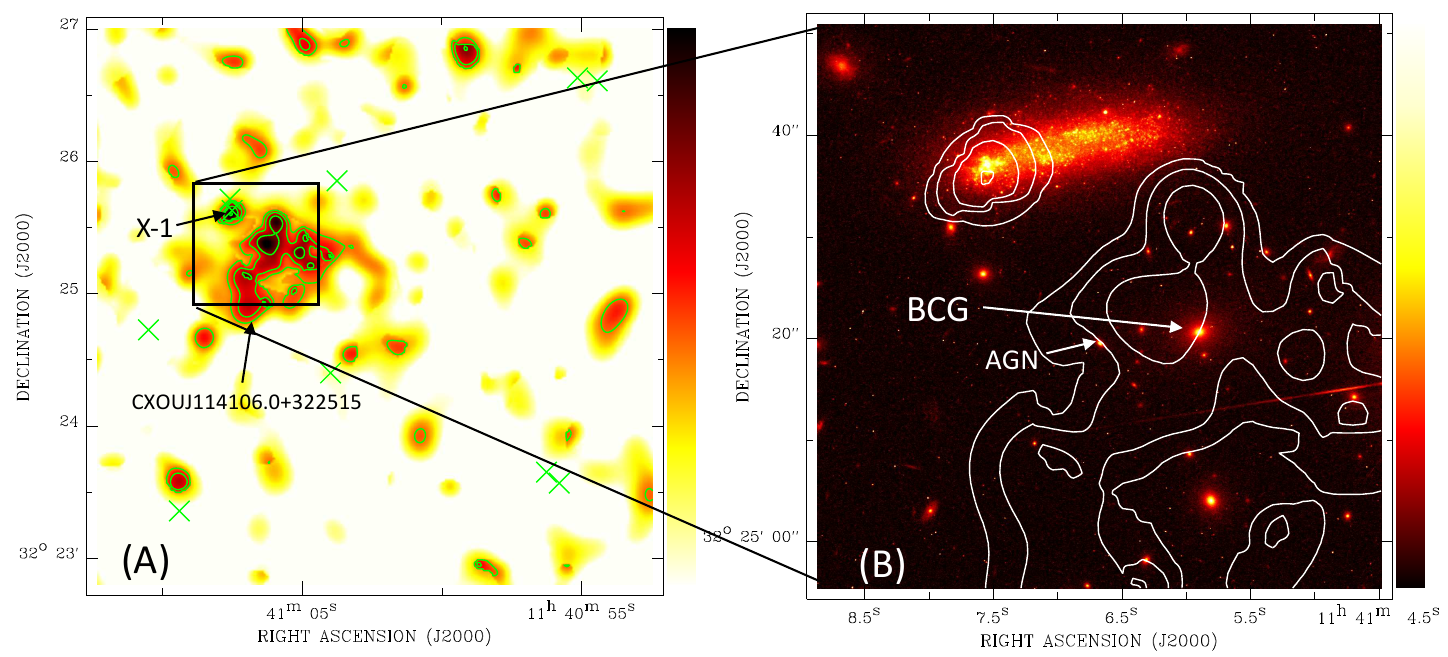}}
\caption{
 {\bf (A)}  Diffuse ACIS-S  0.3-7 keV emission map smoothed with a Gaussian of an adaptive size to achieve a count-to-noise ratio $\gtrsim 5$. The emission intensity contours are at 0.4, 0.5, 0.7, and 1.1 ${\rm~cts~s^{-1}~arcmin^{-2}}$. The positions of removed X-ray sources are marked with "$\times$".  
{\bf (B)} The same intensity contour overlaid on the \hst/WFC3-f814w image. The zoomed-in field of this image is limited by the desire to avoid its outer boundary and the gap between the two detectors. The apparent residual X-ray enhancement at the starburst head of \sou\ is largely due to the PSF wing of \xs\ left from the source removal. Various key objects described in the text are marked.}
\label{f:f3}
\end{figure*}

Fig.~\ref{f:f2} shows the \chandra\ ACIS-S 0.3-7\,keV count image of the \sou\ field. The discrete source, X-1 – the ULX studied in \citet{Wang2025}, stands out. 
The presence of a large-scale extended X-ray emission enhancement is also apparent, as seen in all of the bands. 
Fig.~\ref{f:f3} presents the intensity distribution of diffuse X-ray emission (after excluding discrete sources and smoothing). This emission can be divided into two parts. One is located in and around the star-forming head of the tadpole galaxy and is largely due to the PSF wing of the removed X-1. Any truly diffuse emission of the head is most likely produced locally by the energetic feedback of massive stars. 
The more extended enhancement of the diffuse X-ray emission, called here CXOUJ114106.0+322515, appears south of the galaxy and is very irregular morphologically (Fig.~\ref{f:f3}). No discrete X-ray sources are detected in the enhancement region. 

 For the spectral analysis of CXOUJ114106.0+322515, we extract the on-source and background spectra in the regions defined in Fig.~\ref{f:f2}. With limited counting statistics, we assume that X-ray absorption is dominated by the Galactic foreground H{\sc i} column density fixed at the 21 cm line survey value of $2 \times 10^{20} {\rm~cm^{-2}}$ \citep{Stark1992}. The spectrum is well fitted ($\chi^2/d.o.f.= 28.9/28$) by a simple optically thin thermal plasma with solar metal abundances, temperature $kT = 7.4(5.0-14)$~keV (the 90\% confidence interval), and a 0.3-10 luminosity of log[$L({\rm erg~s^{-1}}$)]=39.78(39.72-39.82) (if at a distance of \sou). We estimate the thermal energy to be $\sim 9\times 10^{56} f_{h}^{1/2} {\rm~erg}$, if the plasma is evenly distributed within a sphere with its radius equal to the spectral extraction radius (0\farcm4 or $2.86$~kpc; Fig.~\ref{f:f2}) and has an effective volume filling factor $f_{h} \sim 1$. Both this energy and the fitted temperature are far too large to be expected of any heating or feedback from the young starburst of the galaxy \citep{Wang2025}. 

\begin{figure*}
\centerline{
\includegraphics[width=1\linewidth,angle=0]{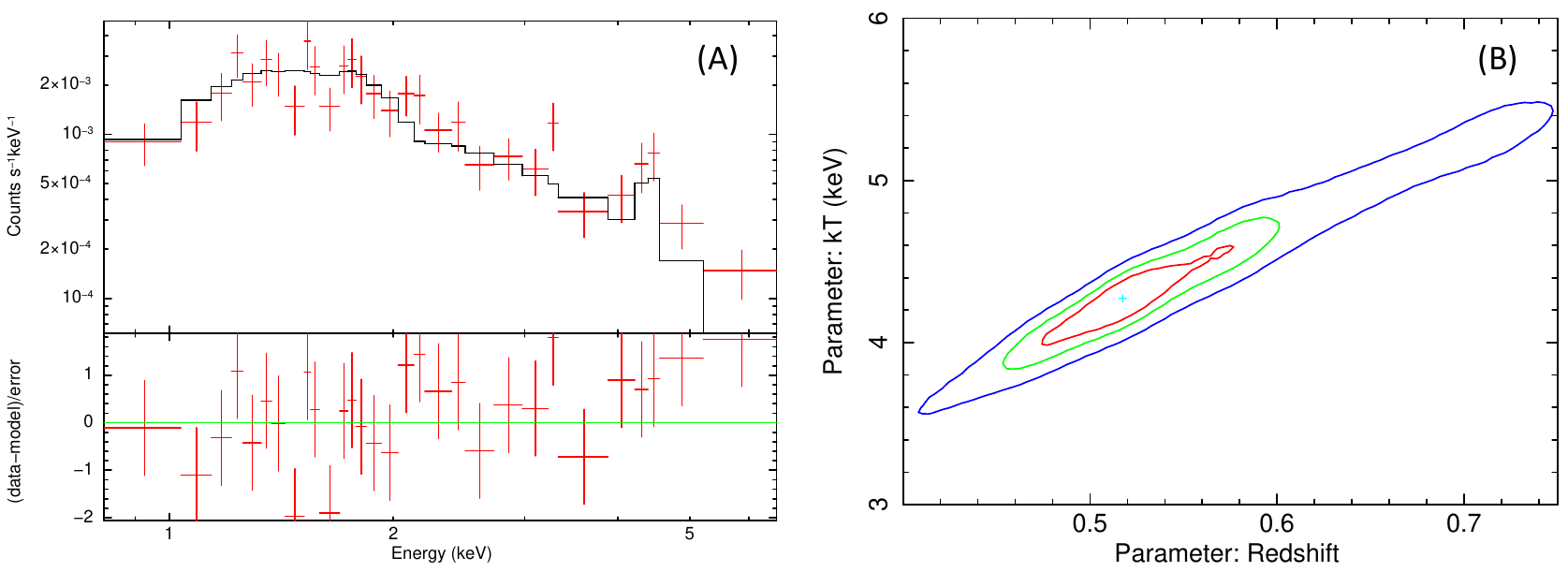}}
\caption{
{\bf (A)} \chandra\ ACIS-S spectrum of the diffuse X-ray emission south of the galaxy \sou, together with the best-fit thermal plasma emission model. {\bf (B)} Confidence contours of the two parameters of the best-fit model at the 68\%, 90\%, and 99\% levels.}
\label{f:f4}
\end{figure*}

\begin{figure*}
\centerline{
\includegraphics[width=1.0\linewidth,angle=0]{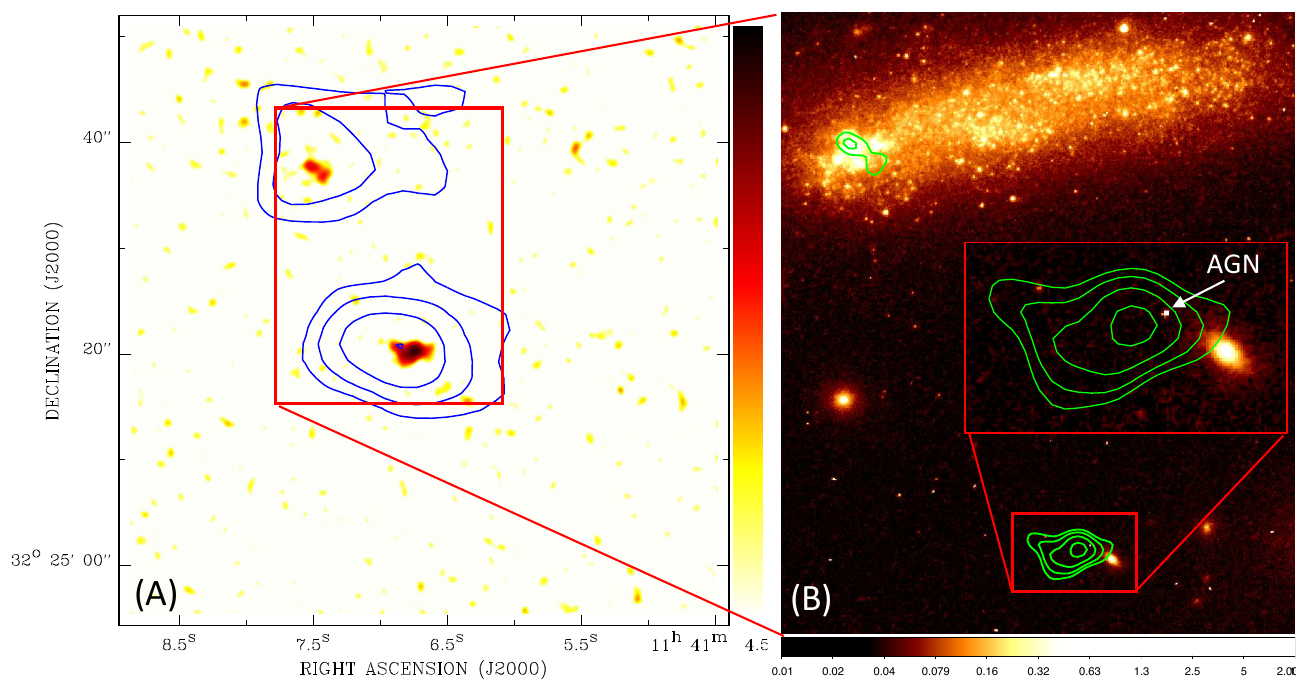}
}\caption{
 {\bf (A)} The VLA 1.5\,GHz continuum intensity map overlaid with the LoFAR intensity contours at 0.25, 0.31, 0.44, and 0.69 ${\rm~mJy~beam^{-1}}$ in the same field as in panel B of Fig.~\ref{f:f3}. The apparent size differences between the VLA and LoFAR features are largely due to their different angular resolutions and sensitivities to diffuse emissions. {\bf (B)} Close-up of the region outlined in the red box (panel A), showing the VLA 1.5\,GHz continuum intensity contours at 60, 80, 110, and 150 $\mu$Jy\,beam$^{-1}$ overlaid on the \hst/WFC3 F814w image. The insert zooms in on the radio lobe associated with CXOUJ114106.0+322515. 
}
\label{f:f5}
\end{figure*}

Instead, we find that the enhancement most likely represents the hot intracluster medium (ICM) associated with a background cluster of galaxies. The enhancement spatially coincides with a concentration of galaxies, preferentially seen in the \hst\ I-band image (Fig.~\ref{f:f3}B). To test this scenario, we fit the spectrum with the model \texttt{TBabs*clumin*APEC} (Fig.~\ref{f:f4}A). To limit the number of fitting parameters, we use the best-fit $L_{x} - kT$ relation of X-ray clusters given in \citet{Giles2016}:
$$L_x = (2.1 \times 10^{43}  {\rm~erg~s^{-1}}) E(z)^{1.64} (kT/3 {\rm~keV})^{2.63},$$
where $L_x$ is in the rest-frame 0.5-2~keV band, while $E(z)$ is a cosmology-related factor and depends on the redshift $z$. The use of the relation allows us to link $L_x$ in \texttt{clumin} to the redshift $z$ and the plasma temperature $kT$ in \texttt{APEC}, where we assume that the plasma metal abundance has a typical value of 0.4 solar. 
The two remaining parameters of the model are well constrained:  $z = 0.52 (0.47 - 0.58)$ and $kT= 4.3 (3.9-4.7)$~keV (see also Fig.~\ref{f:f4}B), while the inferred luminosity of the cluster is $L_{x} \sim 10^{43.5} {\rm~erg~s^{-1}}$. The fit is satisfactory ($\chi^2/d.o.f.= 28.6/28$); in particular, the model explains well the distinct spectral excess observed at $\sim 4.4$\,keV as the Fe 6.7-keV emission line from the rest frame, which is expected from the hot ICM. If fitted, the metal abundance would be 0.78(0.39-1.5) solar, which only marginally improves the fit ($\chi^2/d.o.f.= 25.9/27$), mostly by better matching the Fe line. 

Based on the above results, we consider the X-ray enhancement as the counterpart of the  GMBCG J175.27439+32.42248  cluster identified based on the SDSS data, using the method of a tight cluster red sequence accompanied by a BCG \citep{Hao2010}.  This identification is based on 16 member galaxies, which gives a weighted richness of 16.285. The position of the BCG SDSS J114105.85+322520.9 (or WISEA J114105.87+322520.9; Figs.~\ref{f:f1} and \ref{f:f3}B) is consistent with the centroid of the X-ray feature, while both the spectroscopic redshift of the BCG ($0.5001\pm0.0001$) and the photometric redshift of the cluster (0.492) agree with that constrained by the above X-ray spectral fit.  With the redshift fixed at the BCG value, the fit gives  $kT= 7.4 (5.1-12)$~keV,  which we consider to be our best estimate of the ICM temperature, and log$[L_{x}({\rm~erg~s^{-1}})] = 43.33 (43.25-43.42)$. This luminosity, which is not sensitive to the specific model of the X-ray spectrum, suggests a gravitational mass $M_{500} \approx 2 \times 10^{14} {\rm~M_\odot}$, according to the empirical relation $L_x - M_{500}$ \citep[e.g.,][]{Lovisari2015}. At $z = 0.5001$,  we estimate the total volume (with the assumed spherical shape)  as $V_T=1.4 \times 10^7 {\rm~kpc^3}=4.2 \times 10^{71} {\rm~cm^3}$,  the mean density as $\bar{n}_e \sim 2 \times 10^{-3} {\rm~cm^{-3}}$, and the total thermal energy as $E_T = 3 \times 10^{60} f_{h}^{1/2} {\rm~erg}$, where $f_h$ may effectively account for the inhomogeneity of the hot ICM.  Although not well constrained by the limited counting statistics and energy resolution of the X-ray spectral data, the fitted temperature is considerably higher than $kT= 4.3 (3.9-4.7)$\,keV inferred from the $L_{x} - kT$ relation  \citep{Giles2016}, indicating that the ICM is far from being completely virialized. Furthermore, the X-ray morphology of \xs\ is not close to being centrally peaked or azimuthally symmetric, again suggesting its dynamical youth.  

About $12^{\prime\prime}$  east of the BCG is an extended radio emission feature, observed in both the LoFAR and VLA images (Fig.~ \ref{f:f5}).  
The total fluxes of this feature are $5.0 \pm 1.1$\,mJy and $0.49 \pm 0.12$\,mJy at $\approx 150$\,MHz and 1.5\,GHz. Thus, the inferred spectral index (assuming convention $S_\nu \propto \nu^{-\alpha}$) is $\alpha \approx 1.0 \pm 0.1$. Therefore, this radio emission is nonthermal in origin and presumably arises from the synchrotron emission of relativistic electrons/positrons. West of the feature's peak is a very compact optical source located at $11^h41^m6^s.7,+32^\circ25^{\prime}20^{\prime\prime}$ (see the insert in Fig.~\ref{f:f5}B). The compactness suggests that this previously unclassified object is an AGN. If they are physically associated, then the radio emission may be powered by the jets of this AGN.

\section{Discussion}\label{s:dis}

The above identification and characterization of  CXOUJ114106.0+322515 are essential for any potential study of diffuse X-ray emission associated with \sou. Here, we focus on discussing the implications of our results in understanding the young cluster nature of CXOUJ114106.0+322515 and its relation to the radio lobe.

\subsection{ CXOUJ114106.0+322515 as a young cluster}
 Galaxy cluster assembly is an ongoing process even in the recent universe, as predicted by the hierarchical structure formation model. Clusters at the same redshift can have vastly different assembly histories - some may have formed early and be relaxed systems, while others are still actively assembling through major mergers. Even at low redshifts ($0.2 \lesssim z \lesssim 0.7$), a significant population of galaxy clusters is undergoing mergers with clear evidence from their disturbed X-ray emission morphology and temperature/density substructures that trace shock heating and adiabatic compression of the ICM \citep[e.g.,][]{Markevitch2007, Mann2012,Russell2022, Diwanji2024}. Complementary evidence has also been obtained by spatially resolved observations of the Sunyaev-Zeldovich effect (SZE), which is sensitive to pressure substructure of the ICM, especially at radii larger than those that X-ray observations can readily reach, revealing the extended shock structures and pressure waves generated by mergers \citep[e.g.,][]{Romero2023}, as well as by mapping radio continuum emission, typically at long wavelengths \citep[e.g.,][]{Ignesti2023}.

The X-ray morphological and spectral properties of \xs\ provide useful insights into the structure formation process of a galaxy cluster at $z \sim 0.5$. This cluster shows irregular X-ray morphology, indicating an asymmetric distribution and a lack of a well-defined core of the hot ICM (Fig.~\ref{f:f3}). This morphology appears as complex as the MACS J0717 + 3745 cluster at $z = 0.545$ \citep[e.g.,][]{vanWeeren2017} and the MRC1138-262 protocluster around the well-known Spiderweb Galaxy at $z = 2.16$ \citep[e.g.,][]{Tozzi2022}. The complex morphology of CXOUJ114106.0+322515 thus suggests that it is still being assembled via subcluster mergers and/or collisions of accretion gas streams from the surrounding cosmic web.  One may expect recent intense shock heating, non-equipartition between thermal and bulk motions, or incomplete virialization of the ICM, which may naturally explain its higher temperature estimated from the X-ray spectral fit than that from the $L_{x} - kT$  relation for generally relaxed galaxy clusters \citep{Giles2016}.  Therefore, the cluster is still dynamically young and far from being completely virialized, although its redshift and gravitational mass, as well as the presence of a red sequence, indicate that it has passed the “proton-cluster” phase \citep{Overzier2016}.

\subsection{Energetics of the Radio Lobe and Its Interplay with \xs}
The association of the radio lobe with \xs\ offers a laboratory for studying galaxy feedback during cluster maturation.  How important is the nonthermal component (CR plus magnetic field) of the lobe compared to the thermal hot plasma in the ICM?  Following \citet{Stein2020}, we first estimate the magnetic field strength ($B_{eq}$) in the lobe with the revised equipartition formula by \citet{Beck2005}.  The field in which the synchrotron radiation is emitted is assumed to be completely tangled (no preferred direction).  Adopting the estimated $\alpha \approx 1.0$ and the peak emission intensity $I_\nu \approx 0.2 {\rm~mJy~beam^{-1}}$ measured in the VLA 1.5\,GHz image (Fig.~\ref{f:f5}), we obtain $B_{eq} \approx (29 {\rm~\mu G}) l_{\rm kpc}^{-1/4}$, where $l_{\rm kpc}$ is the characteristic line-of-sight path length (in units of kpc) through the emission region. Noticing the weak dependence  $B_{eq}$ on $l_{\rm kpc}$ and adopting $l_{\rm kpc} \sim 18$, same as the north-south extent of the lobe, we find $B_{eq} \sim 14 {\rm~\mu G}$, greater than the magnetic field strengths ($\sim 1-4$\,$\mu$G) typically estimated in the outer ICM of galaxy clusters or the circumgalactic medium of individual galaxies \citep[e.g.,][]{Govoni2010, Lan2020}. We further estimate the characteristic energetic density and pressure in the lobe of \xs\ as $\epsilon = \epsilon_B + \epsilon_{CR}  = \frac{B_{eq}^2}{4\pi} = 1.6  \times 10^{-11} {\rm~erg~cm^{-3}}$  and $P = (\gamma-1)\epsilon= 5.4 \times 10^{-12}{\rm~dyn~cm^{-2}}$, where the adiabatic index $\gamma = 4/3$ is adopted for the CR and magnetic field. This high pressure  $B_{eq}$ of the radio lobe may be mainly due to its confinement by the ram pressure expected from the fast-moving (or falling) AGN toward the cluster center.  Assuming $P\sim P_\mathrm{ram} = \rho_\mathrm{ICM}v^2$, we estimate the required infall velocity as $\sim 8 \times 10^2 {\rm~km~s^{-1}} [n_e/(2 \times 10^{-3} {\rm~cm^{-3}})]^{-1/2}$, which may be reasonably expected at the offset radius ($\sim 10\farcs6$ or 65\,kpc in projection at $z =0.5$) of the AGN from the BCG. This speed may also result in significant heating of the surrounding ICM \citep[e.g.,][]{Croston2007}, partly explaining its fitted high temperature  (\S~\ref{s:res}).

To estimate the energy involved in producing the radio lobe, we approximate its volume as a prolate ellipsoid. For simplicity, the inclination angle of this ellipsoid (between its major axis and the line of sight) is assumed to be $90^\circ$, which allows us to convert its semi-major and semi-minor axes ($h$ and $r$) of $\sim 1\farcs5$ and 1$^{\prime\prime}$ to 9\,kpc and 6 kpc at a distance of \xs\ and to estimate the volume $V_R =  \frac{4\pi}{3}r^2h \sim 1.4 \times 10^3 {\rm~kpc^3}$. We then infer the total energy as $E = \frac{\gamma}{\gamma-1} P V_R \approx 9 \times 10^{56} {\rm~erg}$, which is roughly a factor of $\sim 3$ greater than the thermal energy of the hot ICM in the same volume (\S~\ref{s:res}).  Therefore, the AGN feedback in the off-center position can indeed play a crucial role in shaping the ICM of the young cluster.

We further constrain the age of the radio lobe.  Our measured spectral slope $\alpha\approx1.0$ is steeper than those ($\approx0.7$) typically observed in the radio lobes around individual galaxies \citep[e.g.,][]{Heesen2014,Zajacek2019} and is approximately 0.5 steeper than the injection index $\sim0.5$ theoretically expected \citep{Bell1978,Blandford1978}. This suggests that the break frequency of the synchrotron emission spectrum has reached $\nu_b \lesssim 0.15$\,GHz -- the effective lower frequency used to estimate $\alpha$. Assuming the above estimate $B_{eq}$, we can then estimate the minimum age of the radio lobe as $\sim 5 \times 10^7$\,years $[\nu_b/({0.15 {\rm~GHz})]^{-1/2}}$, which is comparable to the expected duration of a typical AGN jet ejection episode (probably $\sim 10^7$\,years; \cite{Murgia2011}).  As such, the radio lobe associated with \xs\ may represent a relic of a long-lasting or recent past AGN activity and hence a long-term effect on the structure and dynamics of the ICM. 

\section{Summary}\label{s:sum}

We have studied the diffuse emission around the nearby dwarf galaxy \sou, based on a 90\,ks \chandra\ ACIS-S observation, plus complementary radio data, and we have obtained the following main results and conclusions.
\begin{itemize}
    \item An outstanding diffuse X-ray emission feature (CXOUJ114106.0+322515) is discovered on the southern side of \sou. This feature appears disconnected from the galactic disk and shows a hard X-ray spectrum, which is not expected from a hot plasma outflow from the dwarf galaxy. Our analysis strongly suggests that CXOUJ114106.0+322515 represents a cluster at redshift $\sim 0.5$. This redshift is first indicated in our X-ray spectral fit with a luminosity-temperature relation prior. It is consistent with the optical spectroscopy of an elliptical galaxy located near the feature's centroid and with the photometric redshift of the cluster detected initially via the presence of a red sequence of galaxies. This discovery thus demonstrates the effectiveness of using the prior in estimating the redshift of a galaxy cluster from X-ray CCD data with very limited counting statistics and spectral resolution.  
    \item  The X-ray data have helped to determine the evolutionary stage of the optically identified galaxy cluster, being a dynamical young one. The youth of the cluster is suggested by irregular morphology, lack of a clearly defined center, and unusually high effective plasma temperature of the X-ray emission.  The morphological complexity of the cluster rivals the most extreme cases known for clusters at all redshifts.  We classify \xs\ as a young cluster, although it has apparently passed the proto-cluster phase.
    \item In addition, extended radio emission is detected on the east side of the X-ray feature. This radio lobe has a potential point-like optical counterpart (like an AGN) observed in an \hst\ image. If this association is confirmed, the radio emission is likely to result from the energetic feedback from this AGN. Its offset from the radio emission peak is perhaps due to the ram pressure of the ICM if it is falling toward the cluster's center. The radio emission is nonthermal ($\alpha \approx 1.0 \pm 0.1$) and apparently is of synchrotron origin. 
    \item We estimate the magnetic field strength in the lobe by energy equipartition between the CR and the magnetic field. The estimated high total pressure  $\sim 5 \times 10^{-12}{\rm~dyn~cm^{-2}}$ may be balanced mainly by the ram pressure. The total energy of the radio lobe ($\sim 9 \times 10^{56} {\rm~erg}$) is comparable to the expected thermal energy in the same volume. The synchrotron age of the lobe is $\gtrsim 50$\,Myrs, demonstrating that the CR injection of the AGN likely has long-term effects on the distribution and dynamics of the ICM.  
\end{itemize}

The work reported above underscores the importance of multiwavelength synergy in disentangling overlapping structures (e.g., background young cluster vs. local galactic outflow).  More sensitive observations are desirable to better understand \xs\ and its interaction with the radio lobe. High-angular-resolution radio data (e.g., at high frequencies) can be provided by an upcoming VLA observation of the region, which will enable a detailed morphological and spectral study of the radio lobe. Similarly, high-angular-resolution X-ray data could potentially be provided by the AXIS (Advanced X-ray Imaging Satellite – a NASA concept probe mission) \citep{Mushotzky2019}, allowing spatially resolved X-ray spectroscopy analysis of the hot ICM in this intriguing galaxy cluster. 

\section*{Acknowledgements}
 We thank the anonymous referee for constructive comments that improved the manuscript. 
We acknowledge the support for this work provided by the National Aeronautics and Space Administration via the grant G03-24062X through the Chandra X-ray Observatory Center, which is operated by the Smithsonian Astrophysical Observatory for and on behalf of the National Aeronautics and Space Administration under contract NAS8-03060. We used software provided by the \chandra\ X-ray Center in the CIAO application package. The VLA is part of the National Radio Astronomy Observatory, which is a facility of the National Science Foundation operated under cooperative agreement by Associated Universities, Inc.

\section*{Data Availability}
 This paper employs a list of Chandra data sets, obtained by
the Chandra X-ray Observatory, contained in ~\dataset[DOI: 10.25574/cdc.340]{https://doi.org/10.25574/cdc.340}.

%

\vspace{5mm}
\facilities{Chandra and VLA
}

\bibliography{export-bibtex}{}
\bibliographystyle{aasjournal}

\end{document}